\documentclass{ws-ijgmmp}

\usepackage[utf8]{inputenc}

\begin{document}

\markboth{Authors' Names}
{Instructions for Typing Manuscripts (Paper's Title)}

%
\catchline{}{}{}{}{}
%

\title{REMARKS ON REGULAR BLACK HOLES}

\author{PIERO NICOLINI}

\address{Frankfurt Institute for Advanced Studies (FIAS), Ruth-Moufang-Str. 1\\Frankfurt am Main, D-60438, Germany\\
Institut f\"ur Theoretische Physik, Goethe-Universit\"at Frankfurt am Main, Max-von-Laue-Str. 1\\
Frankfurt am Main, D-60438, Germany\\
\email{nicolini@fias.uni-frankfurt.de} }

\author{ANAIS SMAILAGIC}

\address{Istituto Nazionale di Fisica Nucleare (INFN), Sezione di Trieste,
Strada Costiera 11\\
Trieste, I-34152, Italy\\
smailagic@ts.infn.it }

\author{EURO SPALLUCCI}

\address{Dipartimento di Fisica, Universit\`a degli Studi di Trieste \&
Istituto Nazionale di Fisica Nucleare (INFN), Sezione di Trieste,
Strada Costiera 11\\
Trieste, I-34152, Italy\\
spallucci@ts.infn.it}
\maketitle

\begin{history}
\received{(Day Month Year)}
\revised{(Day Month Year)}
\end{history}

\begin{abstract}
Recently it has been claimed by Chinaglia and Zerbini that the curvature singularity is present even in the so-called regular black hole solutions of the Einstein equations.  In this brief note we show that this criticism is devoid of any physical content.
\end{abstract}

\keywords{Exact solutions; Black holes; Space-time singularities, cosmic censorship, etc..}

\section{Introduction}
In a recent paper~\cite{ChZ17} Chinaglia and Zerbini (CZ) have claimed that a large number of non-singular black hole solutions of the Einstein equations still contain a curvature singularity,  \textit{e.g.} non-linear electrodynamics black holes~\cite{Bar68,AyG98,AyG99,AyG99b,AyG00,AyG05,Ser16}, vacuum filled spacetimes \cite{Dym92,Dym02,Dym03,Hay06}, phantom matter black holes~\cite{BDD03,BrF06,BMD07,BrD07}, regular gravitational shells~\cite{Ans08,AKM87,APS89,FMM89,FMM90,BaP90,NOS13}, noncommutative geometry inspired black holes~\cite{Nic05,NSS06a,NSS06b,Riz06,ANS07,SSN09,Nic09,NiS10,SmS10,MoN10b}, loop quantum black holes~\cite{Mod04,Mod06,Mod08} and
nonlocal gravity black holes~\cite{MMN11,Nic12,BMM14,FZd15,Fro15,FrZ16,FrZ17}.

The CZ reasoning goes as follows. The Einstein equations for a spherically symmetric, static, asymptotically
flat geometry reduce to the single ordinary differential equation: 
\begin{equation}
 r f^\prime + f = 1 -8\pi r^2 \rho \label{eff}
\end{equation}
where, $ g_{00}(r)=-f(r)=-1/g_{rr} $ and $\rho$ is the energy density. CZ claim that the general solution
of this equation is
\begin{equation}
\label{eq:zerbini}
f(r)=1-\frac{C}{r}-\frac{8\pi}{r}\int_0^r du \ u^2\rho(u)
\end{equation}
with $C$ an arbitrary constant. According to them, irrespective of the profile of $\rho$, the metric always exhibits a curvature singularity. This also holds for the  ``\emph{vacuum}'' case, \textit{i.e.},  $\rho\equiv 0$, whose solution would read
\begin{equation}
 f_0(r) = 1-\frac{C}{r}. \label{omo}
\end{equation}
%
Therefore, their claim is that all the papers 
dealing with regular black holes have implicitly assumed $C= 0$ without justification.

\section{Counter-arguments}
\subsection{``Vacuum solution'' and ``solution in vacuum'': massive rabbit out of the magician hat}
There is a very recurrent misunderstanding between ``vacuum solution'' and ``solution in vacuum''. To our knowledge,
 the clearest explanation is given by Richard Feynman in \cite{flecture} referring to  Poisson's equation in electrostatics:
 \begin{quotation}
 Suppose we want a solution of the equation in free space. The Laplacian is
zero because we are assuming that there are no charges anywhere. But the
spherically symmetric solution gives
\begin{equation}
\phi(r)= a + \frac{b}{r}
\end{equation}
in \emph{free} space. \emph{Something is evidently wrong}! The first constant term corresponds
to no charges (empty space) but we have a second term which says that there is a
point charge at the origin. So, although we are solving for a potential in free
space our solution gives the field for a point charge at the origin.
 \end{quotation}
The origin of the misunderstanding is due to the fact that solutions in vacuum are often obtained by solving the Laplace's equation. The mass or the charge (the rabbit) magically resurfaces by invoking the Gauss' theorem. This procedure is customarily used when the source of the Poisson's equation is no longer a function but a distribution.

The lack of distinction between ``vacuum'' and ``in vacuum'' has largely affected the Schwarzschild geometry that is often presented as a vacuum solution, namely
\begin{equation}
 R_{\mu\nu}=0\ ,\quad T_{\mu\nu}\equiv 0\ . \label{vuoto}
\end{equation}
Following the Feynman's reasoning, this conclusion is just wrong. In fact the Scharzschild metric is  a solution outside the source, \textit{i.e.,} in vacuum as 
shown in  \cite{BaN93,BaN94,BaN95,BaN97}.  In other words the Schwarzschild metric is obtained when
\begin{equation}
 \rho= \frac{M}{4\pi r^2}\delta(r) \label{singdens}
\end{equation}
describing a point-like mass $M$ particle at the origin. 
 Thus, following the Feynman's reasoning, the curvature  singularity in $r=0$ is nothing but a ``~warning~'' that something is wrong! The Schwarzschild solution is actually the sum of the ``~\emph{true}~'' 
homogeneous solution  $f=1$ and  the particular solution  $2M/r$ due to the presence of the Dirac-delta source.

The same considerations apply to the solution (\ref{omo}).  It is  \emph{not} a ``vacuum'' solution but a solution in vacuum, \textit{i.e.} it holds everywhere except in $r=0$. 

Apart from mathematical technicalities, the basic results of Einstein gravity is that curved geometries are induced by the presence of matter(energy). Claiming that a singular term is present even in the absence of matter is physically nonsensical.
The only  vacuum solution, compatible with asymptotic flatness, is Minkowski space-time.

\subsection{Simple counter-examples}
For extremely skeptical readers we give two physical counter-examples disproving the CZ line of thought.

\begin{enumerate}
 \item The charged ``shell''.  \\
We start with a general physics textbook example such as the charged empty sphere: 
\begin{equation}
  \nabla^2 \phi = 4\pi \rho,
 \end{equation}
where $\rho \propto \delta(r-R)$. According to CZ, the potential inside the shell would be given as 
\begin{equation}
 \phi= C_0 + \frac{C_1}{r}
\end{equation}
This is clearly nonsense  as there is no electrostatic field inside the shell as predicted by
the Gauss' law.

\item de Sitter metric.\\
Everybody knows that de Sitter space solves 
\begin{equation}
 R_{\mu\nu}=\Lambda g_{\mu\nu}
\end{equation}
On the contrary, following the CZ reasoning, one concludes that the de Sitter metric is
\begin{equation}
 f(r)=1 + \frac{C}{r} -\frac{\Lambda}{3} r^2
\end{equation}
and the curvature diverges in $r=0$. Conversely there is no need to impose $C=0$ since 
the singular term is just not present. 
\end{enumerate}

\section{Conclusions}
We conclude that the statements of the paper
 \cite{ChZ17} do not hold since they are conceptually wrong. CZ simply start from false premises and obtain incorrect conclusions. The false premise is that homogeneous solutions are solutions in vacuum, typically obtained in the case of the Schwarzschild geomety or  potentials generated by point like sources. The incorrect conclusion following such a false premise is that regular sources generate singular solutions. 
 
  As such the CZ work does not shed any light on the problem of regular black holes, nor disproves the existing solutions. 

\label{sec:concl}

\section*{Acknowledgments}

The work of P.N. has been supported by the project ``Evaporation of the microscopic black holes'' of the German Research 
Foundation (DFG) under the grant NI 1282/2-2.

\end{document}